\documentclass[conference]{IEEEtran}

\usepackage{tikz}
\usepackage{amsmath}
\usepackage{amsthm}
\theoremstyle{definition}

\usepackage[utf8]{inputenc}
\usepackage[T1]{fontenc}
\usepackage[english]{babel}
\usepackage{array,graphicx}
\usepackage[squaren]{SIunits}
\usepackage[abbreviations]{foreign}
\usepackage{algorithm}
\usepackage[noend]{algpseudocode}
\usepackage{setspace}
\usepackage{pifont}
\usepackage{listings}
\usepackage{multirow}
\usepackage{subcaption}
\usepackage{stmaryrd}
\usepackage[printonlyused]{acronym}
\usepackage{diagbox}
\usepackage[square,sort,comma,numbers]{natbib}
\usepackage{tabu}
\usepackage{tabularx}
\usepackage{booktabs}
\usepackage{cuted}
\usepackage{paralist}
\usepackage{hyperref}
\usepackage{cleveref}
\usepackage{fixltx2e}
\usepackage{listings}
\usepackage{xcolor}
\usepackage{diagbox}
\usepackage{amsmath}
\definecolor{codegreen}{rgb}{0,0.6,0}
\definecolor{codegray}{rgb}{0.5,0.5,0.5}
\definecolor{codepurple}{rgb}{0.58,0,0.82}
\definecolor{backcolour}{rgb}{0.95,0.95,0.92}
\crefformat{subsection}{\S#2#1#3}
\crefformat{subsubsection}{\S#2#1#3}
\newif\ifshowcomments
\showcommentstrue

\ifshowcomments
\newcommand{\mynote}[2]{\fbox{\bfseries\sffamily\footnotesize{\textbf{#1}}}
 {\small$\blacktriangleright$\textsf{\emph{#2}}$\blacktriangleleft$}}
\else
\newcommand{\mynote}[2]{}
\fi

\newcolumntype{P}[1]{>{\centering\arraybackslash}p{#1}}
\newcolumntype{Y}{>{\centering\arraybackslash}X}

\acrodef{HPA}{host physical address}
\acrodefplural{HPA}{host physical addresses}
\acrodef{HPTW}{hardware page table walker}
\acrodef{PGD}{page global directory}
\acrodef{PML}{Intel page modification logging}
\acrodef{PT}{page table}
\acrodef{PRL}{page reference logging}
\acrodef{PUD}{page upper directory}
\acrodef{VM}{virtual machine}
\acrodef{VMCS}{virtual machine control structure}
\acrodef{WSS}{working set size}
\acrodef{GPA}{guest physical address}
\acrodefplural{GPA}{guest physical addresses}
\acrodef{MMU}{memory management unit}
\acrodef{EPT}{extended page table}
\acrodef{FaaS}{function as a service}
\acrodef{IPI}{inter-processor interrupt}

\def\BibTeX{{\rm B\kern-.05em{\sc i\kern-.025em b}\kern-.08em
    T\kern-.1667em\lower.7ex\hbox{E}\kern-.125emX}}

\title{Out of Hypervisor (OoH): When Nested Virtualization Becomes Practical}
\author{
    \IEEEauthorblockN{Stella Bitchebe}
    \IEEEauthorblockN{Universite Côte D'Azur, ENS Lyon\\
	France\\
    bitchebe@i3s.unice.fr}
    \and
    \IEEEauthorblockN{Alain Tchana}
    \IEEEauthorblockN{ENS Lyon\\ France\\
    alain.tchana@ens-lyon.fr}
}
\begin{document}
\maketitle
%\pagestyle{plain}

%---------------------------abstract---------------------
\begin{abstract}
    This paper introduces Out of Hypervisor (OoH), a new research axis close to nested virtualization. 
    Instead of emulating a full virtual hardware inside a VM to support a hypevisor, OoH principle is to individually expose current hypervisor-oriented hardware virtualization features to the guest OS so that its processes could also take benefit from those features.
    In fact, several hardware virtualization features such as Intel PML, SPP, CAT and EPT which currently can only be used by the hypervisor also be beneficial for processes which run inside the VM.
    We illustrate OoH with Intel PML (Page Modification Logging), a feature which allows efficient dirty page tracking for improving VM live migration.
    According to the fact that dirty page tracking is at the heart of process checkpointing (CRIU) and concurrent garbage collection (Boehm), we present two OoH PML designs namely Shadow PML (SPML) and Extended PML (EPML).
    The former requires no hardware changes but incur significant overhead, justifying EPML which extends PML.
    We evaluated and compared SPML and EPML with \texttt{/proc} and \texttt{userfaultfd}, two default solutions in Linux.
    We do this using a key value store database as the benchmark.
    The results show that EPML reduces CRIU checkpointing time by about 14\% while leading to a negligible overhead (of about 0.5\%) compared to SPML, \texttt{/proc} and \texttt{userfaultfd}.
\end{abstract}    
%---------------------------abstract---------------------

%---------------------------Intro---------------------
\section{Introduction}
\label{introduction}
\begin{figure*}[!ht]
	\centering 
	\includegraphics[width=2\columnwidth]{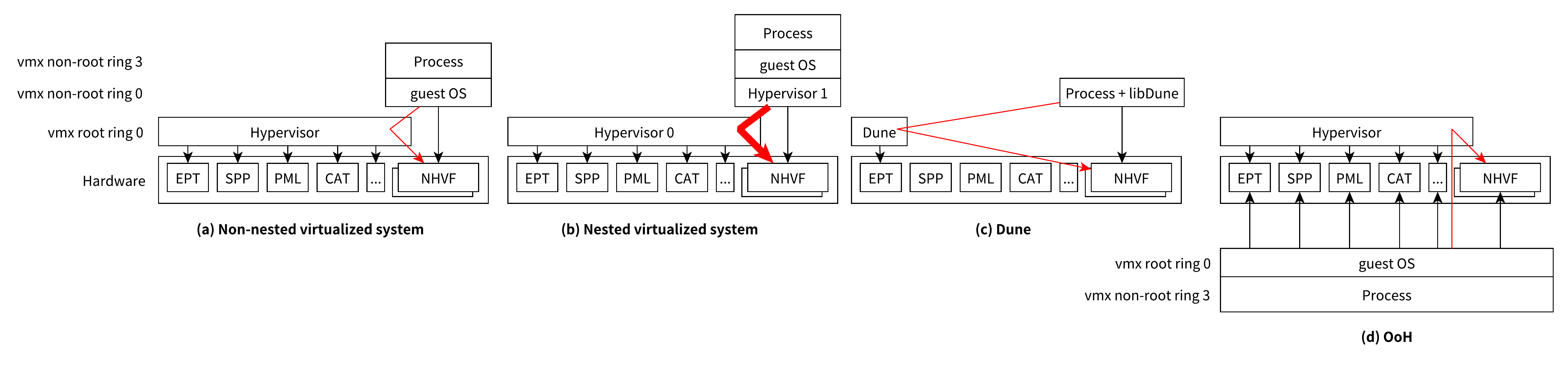}
	\caption{The position of OoH (d) in the virtualization landscape: (a) non-nested virtualization, (b) nested virtualization (including DVH~\cite{lim.dvh.asplos.2020}) and (c) Dune.
	Red arrows materialize VM traps.
	Its width indicates the intensity of VM traps.
	NHVF stands for Non Hardware Virtualization Feature.}
	\label{fig:differences}
\end{figure*}

%\stella{je trouve qu'elle est très longue l'intro non?}
Nested virtualization is the stacking of multiple hypervisors~\cite{lim.dvh.asplos.2020, 
yehuda.turtles.osdi.2010, cloudvisor.zhang.sosp.2011, neve.sosp.2017, vilanova.isca.2019, Goldberg.1974, Belpaire.1975}, see Fig.~\ref{fig:differences}.b.
It induces a significant overhead due to the huge number of VM traps (at least $\times 2$~\cite{vilanova.isca.2019}) compared to non-nested virtualized systems, see Fig.~\ref{fig:differences}.a.
Vilanova et al.~\cite{vilanova.isca.2019} measured that the execution of \texttt{cpuid} instruction is degraded by about 73\% in a nested virtualized system.
Despite the efforts of both the industry and academia to reduce this overhead~\cite{lim.dvh.asplos.2020,yehuda.turtles.osdi.2010,vilanova.isca.2019,neve.sosp.2017}, nested virtualization is recommended by cloud providers such as Microsoft Azure~\cite{azure.nested} for testing, development, and demo.
Even if there is some niche utilization of nested virtualization (to realize rootkits~\cite{cloudskulk.dsn.2021}), its adoption in production is not currently envisioned by cloud users.
Indeed, when cloud customers need isolation and deployability within VMs, they prefer containers (e.g., Docker~\cite{docker}), which are better in terms of performance and have a rich ecosystem (e.g., Kubernetes~\cite{kubernetes}).
Even the ideal VM-based nested virtualization solution would lead to a higher overhead than container-based nested virtualization as containers naturally outperforms VMs.

In this paper we introduce Out of Hypervisor (OoH), a new research axis which advocates that instead of putting a lot of resarch efforts to emulate a full virtual hardware inside the VM for supporting a real hypevisor, we can focus on exposing current \textit{hypervisor-oriented} hardware virtualization features -- such as Intel Page Modification Logging (PML), Sub-Page write Protection (SPP), Cache Allocation Technology (CAT), Extended Page Table (EPT) -- to the guest OS so that its processes (thus containers) can also take benefit from those virtualization features, see Fig.~\ref{fig:differences}.d.
We claim that many hypervisor-oriented hardware features, which currently can \emph{only} be used by the bare metal hypervisor\footnote{The hypervisor which directly runs when the machine is powered-on.} for managing VMs, could also be beneficial for processes which run inside VMs. 
As an example, the exposition to the guest OS of Intel Page Modification Logging (PML), which is currently only used by the hypervisor for dirty 
page tracking for accelerating VM live migration and checkpointing, would allow improving process/container checkpointing (e.g., with CRIU~\cite{criu}) or garbage collection (e.g., with Boehm~\cite{boehm}).
Another example is Intel Sub-Page write Protection\footnote{Intel SPP allows memory protection at the granularity of 128B instead of 4KB.} that would allow reducing memory waste in guard page-based secured heap memory allocators~\cite{10.1145/3133956.3133957,10.5555/3277203.3277213,10.1145/3361525.3361532}.

One may legitimately ask what is the difference between OoH and Dune~\cite{belay.dune.osdi.2012}.
The difference is that \emph{Dune leverages} hypervisor-oriented hardware virtualization features (such as Extended Page Table) to make privileged instructions, which are generally only accessible in kernel mode (ring 0), available to processes, see Fig.~\ref{fig:differences}.c.
\emph{OoH makes} hypervisor-oriented hardware virtualization features accessible to processes in virtualized machines.
%This is more challenging as both the hypervisor and VM's processes should shared (potentially at the same time) the same hardware feature.
Dune processes may take benefit from OoH.

To demonstrate OoH we show in this paper how Intel PML can be exposed to guest's processes.
We call this OoH-PML.
Then we illustrate the power of OoH-PML by using it to speedup CRIU and Boehm GC.
Traditional solutions for dirty page tracking are usualy costly as they are based on write protection, which induces a lot of page faults.
The popular solution under Linux is based on dirty bit and present bit invalidation using the \texttt{/proc} interface.
As shown in~\Cref{sec:motivations}, this solution is extremely expensive and can increase execution time by up to $208.47$\%.
Prior to PML, hypervisors also use such approaches to track dirty pages during VM live migration and checkpointing.
To accelerate these operations, thanks to their importance in the cloud, Intel in collaboration with VMWare introduced PML, a hardware virtualization feature that allows the processor to log in RAM guest physical addresses (GPAs) of pages for which the dirty bit has been set during page table walk.
Xen and KVM, two popular opensource hypervisors, currently support PML when the machine provides it.
Bitchebe et al.~\cite{10.1145/3453933.3454018} empirically validated the benefits of PML in accelerating VM live migration and checkpointing.

PML is currently a hypervisor-oriented feature, meaning that it can only be used by a hypervisor.
However the need for tracking dirty pages is also present within the VM (e.g., process checkpointing or garbage collection).
The application of OoH to PML raises three main challenges:
($C_1$) PML can only be managed by the hypervisor.
We want userspace processes to be able to control PML and to access the PML buffer.
These should be done without hindering the hypervisor functioning. 
($C_2$) PML now works at coarse-grained, that is it concerns the entire VM.
We want to use PML at the granularity of a process inside the VM, while allowing its use by the hypervisor at the scale of the entire VM.
($C_3$) PML only logs GPA.
Userspace processes manipulate guest virtual addresses (GVA).

In this paper, we first explore two designs of OoH-PML.
Second, we show how they can be used by CRIU and Boehm GC.
The first design, that we call Shadow PML (noted SPML), requires no hardware changes as it is based on the current PML implementation.
The second design, that we call Extended PML (noted EPML) extends PML to avoid all the limitations of SPML.
We present SPML in order to justify the necessity of EPML because we know that hardware changes, even minor as in EPML, are not easy to actually implement.
SPML makes the hypervisor emulates PML in software for the guest OS.
Relying on hypercalls and virtual interrupt injection inside the guest, SPML regularly reports logged GPA to a ring buffer shared between the hypervisor and the guest.
It is then up to the guest level to perform reverse mapping from GPA to GVA using the tracked process's page table.
SPML also manages to coordinate hypervisor and guest usages of PML.
Concerning EPML, it hijacks the hardware virtualization features namely VMCS shadowing~\cite{vmcs-shadowing} and posted-interrupt~\cite{posted-interrupt} and slightly extends the original PML to avoid the intervention of the hypervisor when it is used by the guest OS.
VMCS shadowing and posted-interrupt were introduced to allow guest OS directly manipulating virtual machine control structures (VMCS) and receiving interrupt respectively.
We extend PML in two ways.
First, an EPML capable CPU is able to log to two PML buffers simultaneously.
The first PML buffer, that we call guest-level PML buffer, is configured and esclusively controlled by the guest OS.
The second PML buffer, that we call hypervisor-level PML, is configured and esclusively controlled by the hypervisor.
Each tracked thread is associated a guest-level PML buffer that is placed by the guest OS into the processor every time it is scheduled-in.
Second, we extend the CPU to log the GVA (instead of GPA) of the modified page to the guest-level PML buffer.
Note that the GVA is already known by the processor at logging time.

We also present in this paper a library for the utilization of OoH-PML.
That library is independent from the chosen OoH-PML design.
We implement it following the UIO driver principle~\cite{uio}, that is it is composed of two parts: a kernel module and a userspace template code.
The former does not need to be manipulated by the application developper while she should integrate the latter with her application.
We do not rely on the \texttt{/proc} interface, although it would have required no application modification since several applications (e.g., CRIU) already rely on, because it would have required kernel modification.
Also, as \texttt{/proc} is a file system, it brings performance degradation.

We have prototyped and evaluated SPML on a DELL Intel Core i7-8565U processor machine that supports PML, using Xen hypervisor.
Concerning EPML, we rely on BOCHS, the only emulator, to the best of our knowledge, that implements PML.
In both cases, the guest OS is Linux.
We use CRIU and Bohem GC as usecases.
To respect the page limit, we do not discuss Boehm GC results.
The integration of the library requires 251LOC for CRIU.
We use both micro- and macro-benchmarks.
Regarding the latter, we use tkrzw~\cite{tkrzw}, which is a set of database managers.
To evaluate our prototypes, we set up a rigorous methodology, especially for EPML because of the lack of a machine that supports it.
Therefore, we build a formula which quite accurately approximates its performance.

We systematically compare SPML and EPML with two existing dirty page tracking solutions namely \texttt{/proc} and \texttt{userfaultfd}\cite{userfaultfd}.
In terms of performance impact on the tracked process, the solutions are classified as follows, in decreasing order of the overhead they induce:
SPML, \texttt{userfaultfd}, \texttt{/proc}, and EPML.
SPML induces an overhead of up to 6546\% on the micro-benchmark and 58\% on tkrzw Tiny in-memory database engine.
EPML leads to the lowest overhead, which is about 0.5\% on the micro-benchmark and 0.47\% on tkrzw.
Concerning CRIU, SPML lengthens the checkpointing time by about 903\% while EPML reduces it by about 14\%.

In summary, we make the following contributions:
\begin{itemize}
	\item (Principle) We introduce OoH, a new research axis for nested virtualization.
	\item (Empirical contribution) We finely quantify the impact of page fault-based dirty page tracking.
	\item (Conceptual contribution) We present two designs of OoH-PML namely SPML and EPML.
	\item (Technical contribution) We prototype SPML and EPML in real and emulated environments (BOCHS) using popular software systems (Xen hypervisor and Linux guest OS).
	\item (Technical contribution) We integrate OoH-PML into two popular systems namely CRIU and Boehm garbage collection.
	\item (Empirical contribution) We rigourously evaluate our designs using micro- and macro-benchmarks (tkrzw).
\end{itemize}

The rest of the paper is organized as follows.
Section~\ref{sec:background} presents the background.
Section~\ref{sec:motivations} motivates our work.
Sections~\ref{ooh4pml} presents SPML and EPML.
Sections~\ref{sec:usecases} presents the integration into usecases.
Section~\ref{evaluations} presents the evaluation results.
Section~\ref{rw} presents the state of the art.
Section~\ref{conclusion} presents the conclusion.
%---------------------------Intro---------------------

%---------------------------background---------------------
\section{Background}
\label{sec:background}
This section presents the necessary background to understand our contribution.

\subsection{Virtual Machine Control Structure (VMCS)}
\label{sec:background-vmcs}
Intel VT-x (Virtualization Technology extensions) is a hardware virtualization technology for Intel processors.
Its architecture is based on a central design decision that is to not change the semantics of individual instructions of the ISA.
With VT-x, a virtualized environment may allow the guest OS to execute exactely the same instructions as on bare metal.
Intel VT-x introduces two new CPU execution modes namely \texttt{vmx root} and \texttt{vmx non-root}, that are orthogonal to the traditional 4 cpl levels.
The hypervisor runs in \texttt{vmx root cpl 0} mode whereas the guest OS runs in \texttt{vmx non-root mode cpl 0}, thus avoiding the necessity to rewrite the guest OS. 
Interactions and transitions between \texttt{vmx root} and \texttt{vmx non-root} modes are automatically triggered by the CPU using the virtual machine control structure (VMCS) associated to the running vCPU.
In addition to control structures, a VMCS also stores the guest state (when transitioning from guest to hypervisor) and the hypervisor state (when transitioning from the hypervisor to the guest).

VMCS is manipulated using new instructions (e.g., \texttt{vmwrite}, \texttt{vmread}, \texttt{vmlaunch}, etc.).
In its early version, VMCS was only manipulated in \texttt{vmx root} mode.
To improve nested virtualization, Intel recently introduced VMCS shadowing which organizes VMCS in two types: ordinary and shadow VMCS.
A shadow VMCS is a VMCS that is pointed by another one, which becomes an ordinary VMCS.
The latter is only manipulated by the hypervisor while the former can be directly accessed by the guest OS.
Thus in \texttt{vmx non-root} mode, \texttt{vmread} and \texttt{vmwrite} instructions can be executed by the guest OS on shadow VMCS.
Despite the introduction of shadow VMCS, nested virtualization of an entire VM is still inneficient because other hardware features are still unavailable in \texttt{vmx non-root} mode for nested hypervisors.
Our work exploits VMCS shadowing to design an efficient OoH use case.

\subsection{Intel Page Modification Logging (PML)}
\label{sec:background-pml}
PML is a hardware virtualization technology that has been introduced to allow the hypervisor to efficiently monitor guest dirty memory pages.
It relies on Extended Page Table (EPT) and requires specific changes of VMCS. 
When PML is enabled, each write instruction that sets the dirty flag in EPT during page table walk triggers the logging of the GPA at its origin.
To this end, a new 64-bit VM-execution control field called PML Address is added to VMCS.
The PML address points to a \texttt{4KB} memory page, called PML Buffer, that can hold up to 512 logged GPA.
A new 16-bit guest-state field, called PML Index, is also introduced to indicate the index of the next logging entry in PML buffer.
PML index starts at 511.
Whenever the PML buffer is full, the CPU triggers a VMExit and the hypervisor takes the hand.
The handler of that VMExit then copies the content of the PML buffer to a larger buffer (managed by the hypervisor) and resets the PML index to 511.
In Xen and KVM, which are two popular open source hypervisors (used by Amazon EC2), the content of the larger buffer is used to know which pages should be resent during the VM live migration pre-copy phase.
It is important to note that in its current implementation, the activation of PML is machine (concerns all CPUs) and VM (concerns all vCPUs) wide.
In this paper we are trying to make PML efficiently usable from inside a VM, at the granularity of a process.
%---------------------------background---------------------

%---------------------------motiv---------------------
\section{Motivations}
\label{sec:motivations}
\begin{figure*}[!h]
	\centering 
	\includegraphics[width=2\columnwidth]{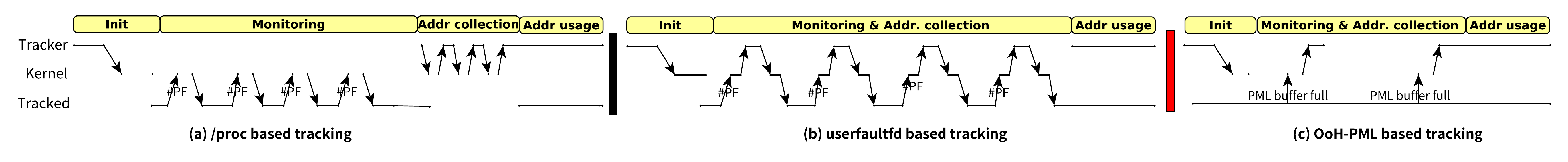}%
	\caption{Impact of \texttt{/proc}, \texttt{userfaultfd} and OoH-PML methods on Tracked and Tracker.
	The two former methods lead to several suspensions of Tracked (due to \#PF).
	\texttt{userfaultd} induces the longest suspension time (\#PF are handled in userspace).
	However, dirty page address collection takes much more time with \texttt{/proc} (due to the parsing of \texttt{/proc/PID/pagemap}), thus impacting Tracker.
	OoH has the benefits of both world and does not require the suspension of Tracked.}
	\label{fig:solutions}
\end{figure*}

Dirty page tracking, thus PML, is not only essential for hypervisors.
A thread which runs inside a VM may also need to monitor dirty pages for garbage collection or checkpointing.
We call Tracker the monitoring thread and Tracked the thread whose memory is monitored.
The traditional approach used by Tracker is the invalidation of dirty and present bits from Tracked's page table entries (PTE).
Linux offers two interfaces that Tracker can leverage.
These are \texttt{userfaultfd} and \texttt{/proc}.
Fig.~\ref{fig:solutions} summarizes their functioning compared to a OoH-PML-based solution.
\texttt{userfaultfd} and \texttt{/proc} are introduced in this section while OoH-PML is presented in~\Cref{ooh4pml}.
The activity of Tracker can generally be organized in four main phases:
the initialization of the tracking method,
the monitoring,
the collection of dirty page addresses and
the exploitation of the latter (e.g., for checkpointing).

We consider in this section that the fourth phase is empty as its duration is agnostic to the tracking method, in comparison with the three other phases.
We launch Tracker and Tracked at the same time but the latter is suspendend during the initialization phase.
The ideal execution time of Tracked is when it runs without beeing tracked.
The ideal execution time of Tracker is the ideal execution time of Tracked.
As one can deduce from Fig.~\ref{fig:solutions}, the choice of the tracking method can impact both Tracker and Tracked.
We can see that OoH is the only method which theoretically leads both systems to their ideal execution time.

\begin{table}[h]
	\scriptsize
	\centering
    \resizebox{\columnwidth}{!}
	{%
        \begin{tabular}{ |l|r|r|r|r|r|r|r| }
            \hline
            On Tracked & 1MB & 10MB & 50MB & 100MB & 250MB & 500MB & 1GB \\ 
            \hline
            \texttt{userfaultfd} & 195 & 272 & 583 & 1,050 & 1,266 & 1,462  & 1,463 \\
            \hline
            \texttt{/proc} & 104 & 55 & 114 & 208 & 302 & 307 & 335\\
            \hline
        \end{tabular}
    }
    \par\smallskip
    \resizebox{\columnwidth}{!}
	{%
        \begin{tabular}{ |l|r|r|r|r|r|r|r| }
            \toprule
            \hline
            On Tracker & 1MB & 10MB & 50MB & 100MB & 250MB & 500MB & 1GB \\ 
            \hline
            \texttt{userfaultfd} & 93 & 169 & 477 & 940 & 1,269 & 1,153 & 1,349 \\
            \hline
            \texttt{/proc} & 47 & 43 & 58 & 148 & 151 & 143 & 147 \\
            \hline
        \end{tabular}
    }
	\caption{Overhead (in \%) of \texttt{userfaultfd}- and \texttt{/proc}-based dirty page tracking methods.}
	\label{tab:basic-clearref-overhead}    
\end{table}

\subsection{The cost of \texttt{userfaultfd}}
Fig.~\ref{fig:solutions}.b summarizes the functioning of a \texttt{userfaultfd}-based dirty page monitoring solution.
To use \texttt{userfaultfd}, Tracker first registers the memory region that it wants to monitor.
After the registration, it will be notified by the kernel each time a page fault which concerns the registered region occurs.
\texttt{userfaultfd} supports two monitoring modes:
\texttt{miss} and \texttt{write\_protect}.
For \texttt{miss}, a notification is sent to Tracker when Tracked accesses a monitored page for the first time.
Concerning \texttt{write\_protect}, a notification is sent when Tracked attempts to modify a monitored page.
In both modes, Tracked is suspended until the fault is resolved.
In the case of \texttt{write\_protect}, the Tracker should write-unprotect the faulted page in order to unpause Tracked.
One can see that, with \texttt{userfaultfd}, the collection of dirty page addresses can be done during the monitoring phase.

We assess the overhead of \texttt{userfaultfd} using as Tracked a synthetic program (presented in~\ref{evaluations}) that just parses and writes to an array of buffers.
The size of each buffer is 4K, allocated at page boundaries.
We are interested in monitoring the entire array.
Table~\ref{tab:basic-clearref-overhead} second and fifth rows show the overhead of \texttt{userfaultd} while we vary the array size.
We can see that the overhead linearly increases with the array size.
We measured an overhead of up to 1,462\% and 1,349\% for 1GB on Tracked and Tracker respectively.
We breakdown the page fault handling time into two components:
the time spent inside the kernel (about 33.6ms for 1GB) and the time spent in Tracker (about 3,3383ms for 1GB).
The total suspension time of Tracked represents in average about 93\% of its execution time.

\subsection{The cost of \texttt{/proc}}
Fig.~\ref{fig:solutions}.a summarizes the functioning of a \texttt{/proc}-based dirty page monitoring solution.
Tracker first instructs the kernel to clear soft-dirty bits of Tracked's PTEs.
This is done by writing 4 to \texttt{/proc/PID/clear\_refs} file, where PID is the process identifier of Tracked.
This operation is dominated by the time taken by the kernel to parse Tracked's PTEs and to flush the TLB (about $2.234 ms$ when the monitored memory is 1GB).
All of this lengthens the intinitialization step compared to \texttt{/userfaultfd}.
After this, once Tracked tries to modify a monitored page, a fault occurs.
The handler of that fault sets in Tracked's PTE the soft-dirty bit of the faulted page (this operation costs about $33.5 \mu s$). 
At the end of the monitoring period, Tracker reads the soft-dirty bits (bit 55) from \texttt{/proc/PID/pagemap} in order to determine all dirty page addresses (this costs about $594.187 ms$ when the monitored memory is 1GB). 
The total suspension time represents about 73\% of the total execution time of Tracked.
As shown in table~\ref{tab:basic-clearref-overhead} top, the impact of \texttt{/proc} on Tracked varies with the memory size. 
We measured an overhead of about $335.24$\% for 1GB of memory. 
This overhead is lower than the one induced by \texttt{/userfault} as shown above.
Concerning Tracker, see table~\ref{tab:basic-clearref-overhead} bottom, the overhead varies from 47\% up to 147\%.
When compared to \texttt{/userfault}, algthough \texttt{/proc} increases the address collection phase, its cost is compensated by the smaller suspension it induces during the monitoring phase.
%---------------------------motiv---------------------

%---------------------------contrib---------------------
\section{OoH-PML: Design and Implementation}
\label{ooh4pml}
Althought our design is generic, we rely on a Xen virtualized system to describe it.
The guest OS is Linux.
The goal of OoH-PML is to make PML usable from inside the guest OS.
We do this with as minimum as possible changes in the hardware, the hypervisor, the guest kernel.
In addition, we want to propose a simple utilization interface to application developers.

%---------------------------contrib-overview---------------------
\subsection{General overview}
\label{overview}
We study two designs namely Shadow PML (noted SPML) and Extended PML (EPML).
SPML requires no hardware modification while EPML introduces a slight extension of the original PML feature.
SPML exploits all existing mechanisms to make an OoH version of PML that works as efficiently as possible with existing machines.
Despite these efforts, we measured a very significant overhead due to the current architecture of PML, thus justifying EPML.
The latter leverages existing nested virtualization features (such as shadow VMCS and Posted-Interrupts) and requires minor hardware changes in order to provide the ideal OoH-PML performance.

The utilization of OoH-PML requires a small modification of Tracker.
In fact, we provide OoH-PML to userspace following the UIO driver principle, that is the driver is organized in two components:
a kernel module and a userspace code.
This organization brings flexibility in our design that is the kernel and the implementation of OoH-PML can evolve independently.
The userspace code that we provide is a template that the Tracker developer should integration.
This template includes the following steps.
First the code registers with the UIO kernel module the PID of Tracked.
This allows the kernel to mark Tracked as to be monitored using OoH-PML.
After that, the template code mmaps a ring buffer (see below its initialization) and loops on it.
This buffer is the place where the addresses of dirty pages are stored by OoH-PML.
Contrary to the original PML which logs GPA, OoH-PML logs GVA.
At the end of the loop, Tracker can unregister Tracked with the kernel to stop its monitoring.

The UIO kernel module is at the heart of the software part of OoH-PML.
At load time, it does some initialization operations such as:
hardware initialization of PML,
allocation and initialization of the ring buffer which is shares with the hypervisor (for SPML only) and future Trackers.
When a Tracker registers with the module for monitoring, the latter installs with Linux scheduler two callbacks leveraging the prempt-notifier interface: one for schedule-in events and the second for schedule-out events.
This way, the module can know when to enable or disable PML.

Although the utilization of the UIO driver principles requires a slight modification of Tracker, we choose this approach rather than \texttt{/proc} which is likely already taken into account several Tracker applications (such as CRIU) for two main reasons: performance and code scalability.
First \texttt{/proc} is a file system, thus its access requires to walk through the file system stack which is known to degrade performance.
Second, to integrate \texttt{/proc} with OoH-PML, we should modify the core of the guest kernel (Linux) for two purposes:
(1) implementation of a new version of \texttt{echo 4 > /proc/clear\_refs};
(2) for each PML buffer full event, the handler should walk through Tracked's PTE to set the dirty bit for the corresponding entries. 
We showed in~\Cref{sec:motivations} that theses operations are costly.
%---------------------------contrib-overview---------------------

%---------------------------contrib-desc---------------------
\subsection{Shadow PML (SPML)}
\label{sec:shadow-pml}
The basic idea behind SPML is to make the hypervisor emulates PML for the guest.

\subsubsection{PML initialization and logging}
\label{sec:shadow-pml-init}

The actual initialization of PML is done by the hypervisor when loading the guest loads the UIO kernel module.
It has no effect if PML was already initialized by another VM.
It consists in initializing VMCS structures, allocating a region in RAM for the PML buffer and initializing the PML index.
As the intiatialization of PML concerns all vCPUs of the VM, processors may log to the same buffer GPA from any process runing inside the VM.
This raises the first challenge ($C_1$) of SPML as, we aim to log only the addresses of processes that the user wants to track in the guest. 
Indeed, if the guest kernel schedules-in a process that does not need to be tracked, the processor will log to the PML buffer the addresses generated by that process, which are not those needed by Tracker.
To alleviate this, we introduce two new operations for \texttt{HYPERVISOR\_vcpu\_op} hypercall: \texttt{enable\_logging} and \texttt{disable\_logging}. 
We use these hypercalls to control the processor logging process at low cost\footnote{In fact, our PML \texttt{enable\_logging} is different from the PML intialization operation which is more costly.} as follows.
When a tracked process is scheduled-out, the guest kernel performs a \texttt{disable\_logging} hypercall and the hypervisor retrieves from the PML buffer all the addresses that have been logged and sets the PML index to $512$ to prevent logging.
When a tracked process is scheduled-in, the guest kernel performs a \texttt{enable\_logging} hypercall and the hypervisor resets to $511$ the PML index to make the CPU restart its logging activity.

\subsubsection{PML buffer handling}
\label{sec:shadow-pml-handle}
When the PML buffer is full (at the hypervisor level), the processor stops logging and triggers a vmexit so that the hypervisor can retrieve the addresses and reset the PML index. 
In SPML we set up a ring buffer to which the hypervisor copies the logs.
When the ring buffer itself is full, the hypervisor triggers a virtual interrupt at the destination of the guest OS, which is handled by our UIO kernel module.
The latter in turn notifies Tracker in userspace.
No more address is filled in the ring buffer until it is flushed.
Notice that, Tracker can manage to regularly collect addresses from the ring buffer in order to avoid this situation.
This is what we do in our experiments.

Because there is a unique PML buffer per vCPU, when many processes are tracked simultaneously we need to ensure the homogeneity of logs while filling the ring buffer.
This is the second challenge ($C_2$) of SPML: how to identify logged addresses per process?
To answer this question, we organize the ring buffer in blocks in which the two first entries are respectively the PID  of a tracked process (filled in by the UIO driver module when scheduling the process in) and the current number of addresses logged for this process (incremented by the hypervisor when filling the ring buffer and decremented by Tracker when collecting logs).

\subsubsection{Coexistence with the hypervisor}
\label{sec:shadow-pml-coexistence}
SPML must not interfere with the utiliation of PML by the hypeprvisor.
This is the third challenge ($C_3$).
Let us imagine that the guest asks for deactivating PML while the hypervisor is still using it (e.g., to perform live migration) or vice versa. 
We coordinate their intervention as follows.
(1) We add two flags \texttt{enable\_by\_vmm} and \texttt{enable\_by\_guest} per vCPU to indicate which level has initialized PML.
(2) We add some control checks in the PML buffer shuching code path inside the hypervisor so that:
if \texttt{enable\_by\_vmm} is not set, the hypervisor does not execute the original portion of code corresponding to its own use (this may incur unnecessary additional CPU time utilization as the hypervisor is not actually using PML);
if \texttt{enable\_by\_guest} is not set, the hypervisor does not fill the ring buffer.
(3) When the guest asks to deactivate PML, a control check first verifies that \texttt{enable\_by\_vmm} is not set (meaning that the guest was the only level to use PML) and vice vesa.
If the hypervisor or the guest asks to disable PML while the other is still using it, we just unset the corresponding flag without effectively disabling PML on the processor.
(4) On schedule-out of a tracked process, as stated in \Cref{sec:shadow-pml-init}, the guest instructs the hypervisor to stop logging addresses for its use.
In this case, the hypervisor first flushes the PML buffer so that each of the hypervisor and the guest may collect addresses that have been logged till this moment, and if \texttt{enable\_by\_vmm} is set, we just unset another flag \texttt{sched\_in} that tells the hypervisor if a tracked process is scheduled-in (and thus if it must fill the ring buffer or not);
otherwise (i.e., the hypervisor is not using PML), we set the PML index to $512$.
Doing so, the CPU will stop logging, which avoids unnecessary CPU time waste and additional vmexits as neither the hypervisor nor the guest needs logs at that moment.
(5) On schedule-in of a tracked process, the guest also instructs the hypervisor to restart collecting addresses and filling the shared ring buffer.
In this case, we need to flush the PML buffer to avoid collecting addresses that do not belong to the current tracked process.
But before doing this, the hypervisor checks the flag \texttt{enable\_by\_vmm}: if set, the hypervisor first copies the addresses that are in the PML buffer and resets the PML index. 

To verify that this coordination protocol effectively works, we performed five iterations of live migrations on a VM using SPML.
We observed no crash of the VM and no interruption of live migration.

\subsubsection{Address collection in userspace}
\label{sec:shadow-pml-coexistence}
With PML, the processor logs the GPA of the modified pages while userspace needs GVA.
This is the fourth challenge ($C_4$) of SPML.
To address this challenge, the template code released with SPML includes a portion of code which performs a \emph{reverse mapping} operation to translate each GPA to GVA every time it reads the ring buffer in the loop. 
To do so, it walks through the tracked process's page table relying on \texttt{/proc} interface.

\subsubsection{SPML limitations}
\label{sec:shadow-pml-limit}
SPML leads to two limitation types: performance overhead and address lost.
Performance overhead is due to two factors: a potential huge number of hypercalls (at each process schedule-in and out event) and the reverse mapping operation to translate GPA to GVA.
The former factor impacts both Tracked and Tracker while the latter factor only impact Tracker.
Concerning address lost, if a GPA is no longer present in the page table (because of an unallocation) of the tracked process at reverse mappin time or has been assigned to another GVA, the corresponding GVA would either be nill or inaccurate.
All these limitations motivate Extended PML.

\subsection{Extended PML (EPML)}
\label{sec:extended-pml}

The basic idea behind EPML is to provide a second level of PML that is entirely controlled by the guest OS.

\subsubsection{PML initialization and logging}
\label{sec:extended-pml-init}
\begin{figure}[!h]
	\centering 
	\includegraphics[width=.8\columnwidth]{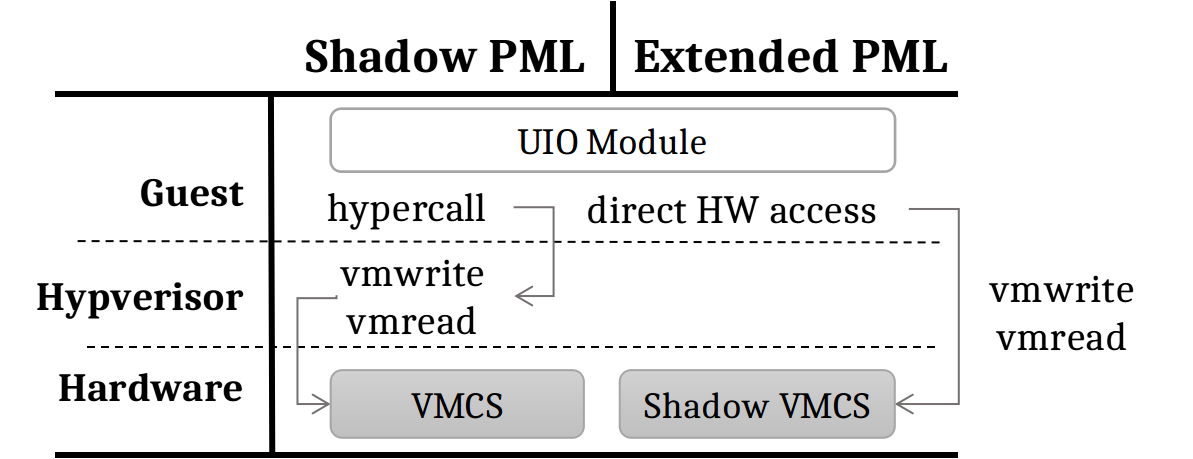}
	\caption{Hardware access difference between EPML and SPML.}
	\label{fig:designs-diff-access}
\end{figure}

From a software point of view, the differences between EPML and SPML are as follows (see Fig.~\ref{fig:designs-diff-access}):
(1) the ring buffer is no longer shared with the hypervisor;
(2) the flags \texttt{enable\_by\_vmm} and \texttt{enable\_by\_guest} are no more used.
To minimize hardware changes, EPML leverages the existing VMCS shadowing feature.
At kernel module load time, the hypervisor is called to initialize the shadow VMCS pointed to by each of ordinary VMCS maintained for the VM.
This is the only hypercall performed in EPML. 
This initialization consists of allocating a memory region for the shadow VMCS and filling its address in the VMCS link pointer field of the ordinary VMCS, and initializing the fields \texttt{VMREAD\_BITMAP} and \texttt{VMWRITE\_BITMAP} of the shadow VMCS. 
These bitmaps allow the hypervisor to specify which fields of the shadow VMCS the guest is allowed to read or write without trapping.
When a tracked process is scheduled-in or out, the kernel accordingly enables or disables the logging process by setting the PML index to $512$ in the shadow VMCS using a \texttt{vmwrite} instruction. 

From a hardware point of view, EPML makes the following changes.
We introduce in the VM-Execution Control area of VMCS a new field (called \texttt{Guest PML Address}) that represents the address of the memory region where the processor will log the addresses for the use of the guest OS.
EPML allows the guest to configure a guest-level PML buffer per tracked process, exactely as the hypervisor manages a PML buffer per vCPU. 
Subsequently, when a new process is registered to be tracked, the guest kernel allocates a memory region for its guest-level PML buffer. 
Further, when this process is scheduled-in, before enabling the logging process, the guest kernel loads its corresponding guest-level PML buffer by filling its address in the Guest PML Address field of the shadow VMCS using a \texttt{vmwrite} instruction. 
Because the guest only sees GPA, the value that it sets to the Guest PML Address field should be translated to a host physical address (HPA) so that the processor can log to the right location in RAM.
To tackle this, we extend the VMX ISA so that if a \texttt{vmwrite} instruction is performed (to fill the Guest PML Address field) when the processor is in VMX non-root mode, it first translates the address to a HPA (using EPT or TLB) before writing it to the shadow VMCS.
We claim that this translation is very less expensive than performing a hypercall as in SPML.

Another improvement brought by EPML is the capability of the processor to log GVA in order to avoid reverse mapping of GPA as SPML requires.
We modify the page walk process so that the processor can log the GVA to the guest-level PML buffer and the GPA to the hypervisor-level PML buffer.

\subsubsection{Buffer full event handling}
\label{sec:shadow-pml-handle}
\begin{figure}[!h]
	\centering 
	\includegraphics[width=1\columnwidth]{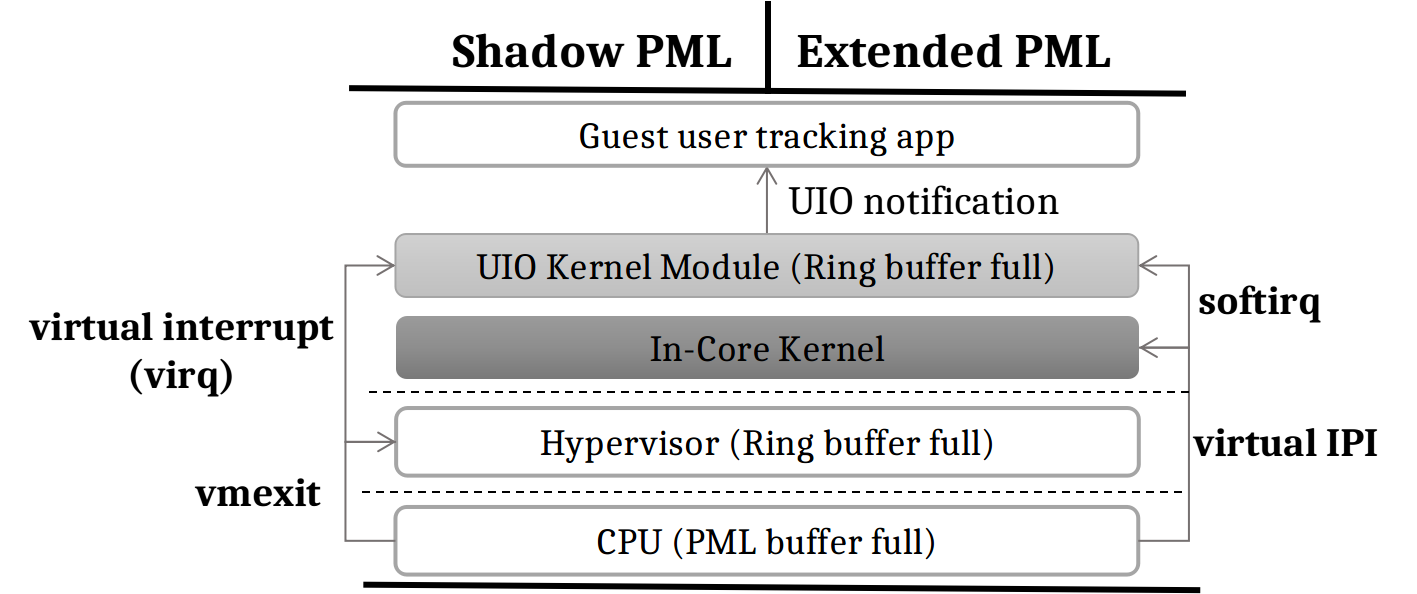}%
	\caption{Buffer full handling difference between EPML and SPML.}
	\label{fig:full-handle-diff}
\end{figure}

Fig.~\ref{fig:full-handle-diff} summarizes the difference between EPML and SPML.
To handle guest-level PML buffer full events, EPML leverages Posted-Interrupts.
With the latter the processor is able to deliver a virtual interrupts in vmx non-root mode without vmexits.
Using this functionality, we modify the hardware behavior so that when a guest-level PML buffer is full, the processor raises a virtual \emph{self-IPI (Inter-Processor Interrupt)}.
This interrupt is handled by the UIO kernel module.
As the latter services a \emph{virtual} device, it cannot directly receive an interrupt from the processor. 
So we define a new interrupt vector in Linux that the processor uses for the virtual IPI.
To redirect the handling of that virtual IPI to the UIO module, we rely on the Linux-Softirq principle. 
We define a new softirq and divide the handling of the virtual interrupt into two parts:
the top-half in the core kernel and the bottom half in the UIO module. 
The topf half just raises the softirq while the bottom half copies the content of the guest-level PML buffer to the ring buffer.

\subsubsection{Coexistence with the Hypervisor}
\label{sec:shadow-pml-coexistence}
VMCS shadowing was introduced for the purpose of nested virtualization.
Once initialized by the hypervisor, the latter no longer interfers with its utilization.
Anyway, as the VM and the hypervisor could be both using PML simultaneously we must just pay little attention on PML index. 
Along with the guest-level PML buffer address, we also introduce in the VMCS structure a new field for its PML Index.
%---------------------------contrib-desc---------------------

%---------------------------contrib---------------------

%---------------------------implem---------------------
\section{Use case}
\label{sec:usecases}
We study the utilization of OoH by CRIU and Boehm GC, which only consists in integrating the userspace template code with these systems.
Due to page limit, we only present CRIU in the present paper.

\paragraph{CRIU overview}
CRIU is a Linux software that can checkpoint to disk the state of a process or a container.
It is currently integrated into many well-knwon projects such as OpenVZ~\cite{openvz}, Podman~\cite{podman} or Docker~\cite{docker}.
To checkpoint a process, CRIU saves the state of the latter as a collection of files: we talk about \emph{dumping} the application.
CRIU supports two dumping modes namely incremental and non-incremental. 
Non-incremental dumping consists in saving the entire memory state of the process while in incremental mode CRIU only dumps dirty pages since the previous dump. 
Thus, OoH could be beneficial for incremental dumping.
In respect with our nomenclature, CRIU is a Tracker.
In its current implementation, it relies on \texttt{/proc}.
To integrate our solution, we mainly patched two CRIU steps (the reader shoud remember the four steps of Tracker introduced in~\ref{sec:motivations} and Fig.~\ref{fig:solutions}):
\begin{itemize}
	\item Initialization phase:
	In respect with the utilization \texttt{/proc}, CRIU first pauses the tracked tasks before initialization (\texttt{echo 4 > /proc/PID/clear\_refs}), as illustrated in Fig.~\ref{fig:solutions}.a.
	With OoH, this pause is no longer necessary as the activation of PML is immediate and does not interfer with the execution of the tracked thread, as illustrated in Fig.~\ref{fig:solutions}.c.
	\item Address collection:
	We discard from CRIU the portion of code which parses \texttt{/proc/PID/pagemap} for collecting addresses as with OoH addresses are collected during the monitoring phase.
\end{itemize}
The patch adds 251 LOC to CRIU in 9 files.
%---------------------------implem---------------------

%---------------------------evals---------------------
\section{Evaluations}
\label{evaluations}
The goal of the evaluation is to answer the following questions:
(1) what is the potential overhead or improvement of SPML and EPML compared to existing solutions (\texttt{/proc} and \texttt{userfaultfd})?
(2) what is the scalability of SPML and EPML?
(3) to what extend the utilization of PML from inside the guest impacts its utilization by the hypervisor?
(4) to what extend SPML and EPML are able to capture the entire dirty page working set?

\subsection{Experimental environment}
\label{environment}

\paragraph{Machine and Systems}
We carried out our experiments on a machine with 8 Intel Core i7-8565U and 16 GB of RAM. 
The virtualization system we use is Xen 4.10.0 and each VM runs Ubuntu 18.04 with Linux 4.15.0. 
Especially to evaluate \texttt{userfaultfd} we use Linux 5.11 because version 4.15.0 does not support the \texttt{write\_protect} mode.
Concerning the emulation of EPML, we use BOCHS 2.6.11.
In all experiments on the real machine, the VM has 4 dedicated vCPUS with 5GB of memory. 
In the emulated environment, the VM has 1 vCPU and 1GB of memory due to BOCHS constraints. 

\paragraph{Benchmarks}
We use a micro- and a macro-benchmark.
The former is a synthetic algorithm which parses an aray of 4KB buffers, see listing~\ref{lst:microbench}.
Its performance metric is the execution time.
For the macro-benchmark, we use tkrzw~\cite{tkrzw}, which is a suite of database managers.
We evaluated all the five in-memory engines that tkrzw provides using a load injector which performs \texttt{set} requests.
Table~\ref{tab:tkrw-configuration} shows for each manager the configuration setup.

\begin{table}[h]
	\scriptsize
	\centering
	\resizebox{\columnwidth}{!}
	{%
		\begin{tabular}{ |l|c|c|c|c|c|c| }
			\hline
			\backslashbox[30mm]{Parameter}{Engine}  & Baby 		& Cache 	& Stdhash 	& Stdtree 	& Tiny \\ 
			\toprule
			\#threads 			& 3 		& 5 		& 4 		& 7 		& 5  \\        
			\hline
			\#iterations 			& 10M	& 10M	& 10M 	& 100M & 10M \\
			\hline
			\#max records 		& - 		& 10M 	& - 		& - 		& - \\
			\hline
			compression mode 	& - 		& - 		& zlib 		& - 		& - \\
			\hline
			\#buckets for hashing & - 		& - 		& 10M 	& - 		& 100M  \\
			\hline
		\end{tabular}
	}
	\caption{The configuration set-up for each tkrzw engines. ($M$ means million)}
	\label{tab:tkrw-configuration}    
\end{table}

\begin{figure}[thp]
\centering
\begin{tabular}{c}
\begin{lstlisting}[language=C, caption={The micro-benchmark code.}, aboveskip=\medskipamount, label={lst:microbench}, basicstyle=\scriptsize]
... 
#define PAGE_SIZE sysconf(_SC_PAGE_SIZE)
#define num_pg xx //memory size=xx*PAGE_SIZE

void main(void)
{
	unsigned long *region=malloc(num_pg*PAGE_SIZE);
	/*
	 * Pin all the pages in-memory to be sure that 
	 * they are not swapped out
	 */
	mlockall(MCL_CURRENT|MCL_FUTURE|MCL_ONFAULT); 
	for( ; ; )
		for(unsigned long i=0;i<num_pg;i++)
			region[(i*PAGE_SIZE)/sizeof(unsigned long)]=i;
}
\end{lstlisting}
\end{tabular}
\end{figure}

\begin{table*}[!htb]
\scriptsize
\center
		\begin{tabular}{ | l | c | c | r | } 
			\hline
			\textbf{Metric} 							& \textbf{Depend on mem.} & \textbf{Cost (\boldmath$\mu$s)} & \textbf{Technique} \\ 
			\toprule
			\hline
			\textbf{$M1$.}  context switch (from user to kernel space)	& 	No						   & $0.315$							  &  All \\ 
			\hline
			ioctl syscalls  					& 							   & 									  &   \\ 
			\textbf{$M2$.}  write\_protect 					& 	Yes						   &  - 								  & userfaultfd  \\  
			\textbf{$M3$.}  init. PML 						& 	No						   &  $5,651$						  & SPML \& EPML  \\
			\textbf{$M4$.}  deactivate PML 						& 	No						   &  $2,816$						  & SPML \& EPML  \\
			\hline
			page fault handling  		& & & \\
			\textbf{$M5$.}  in kernel space 						&  	Yes						   &  -				 					  &  /proc, userfaultfd \\
			\textbf{$M6$.}  in userspace	&  	Yes						   &  -									  &  userfaultfd \\
			\hline
			vmx operations 					& & & \\
			\textbf{$M7$.}  vmread 								&  	No						   &  $0.936$							  &  EPML \\
			\textbf{$M8$.}  vmwrite 								&  	No						   &  $0.801$							  &  EPML \\ 
			\hline
			hypercalls  						& & & \\ 
			\textbf{$M9$.}  init. PML 						& No						   &  $5,495$						  &  SPML\\ 
			\textbf{$M_{10}$.}  \qquad  + init. VMCS shadowing	& No						   &  $5,878$						  & EPML \\ 
			\textbf{$M_{11}$.}  PML deactivation 					& No						   &  $2,060$						  &  SPML \\ 
			\textbf{$M_{12}$.}  \qquad  + VMCS shadowing deactivation & No						   &  $2,755$						  & EPML \\ 
			\textbf{$M_{13}$.}  enable PML logging 					& No						   &  $0.3$									  &  SPML \\
			\textbf{$M_{14}$.}  disable PML logging 					& Yes						   &  -									  &  SPML \\
			\hline
			\textbf{$M_{15}$.} \texttt{echo 4 > \texttt{/proc/PID/clear\_refs}}	&	Yes				   &  -									  &  /proc \\ 
			\hline
			\textbf{$M_{16}$.} page table walk (in userspace)	& Yes						   &  -									  &  /proc \& SPML \\ 
			\hline
			\textbf{$M_{17}$.} reverse mapping  					& Yes						   &  -									  &  SPML \\ 
			\hline
			\textbf{$M_{18}$.} ring buffer copy  					& Yes						   &  -									  & EPML \& SPML \\ 
			\hline
		\end{tabular}
	\caption{Cost of internal metrics which are agnostic to Tracked memory size.}
	\label{tab:basic-metrics-1}		
\end{table*}

\begin{table*}[!htb]
\scriptsize
\center
			\begin{tabular}{|l||*{7}{c|}}\hline
				\backslashbox[]{Metric (time in ms)}{Memory size} 	&	\makebox[3em]{1MB}	&	\makebox[3em]{10MB}	&	\makebox[3em]{50MB}	&	\makebox[3em]{100MB}	&	\makebox[3em]{250MB}	&	\makebox[3em]{500MB}	&	\makebox[3em]{1GB}	\\
				\toprule\hline
				\texttt{echo 4 > /proc/PID/clear\_refs}		& 	0.032	&	0.0912	& 	0.174		& 	0.288		& 	0.613		& 	1.153		& 	2.234	\\
				\hline
				Page table walk	in userspace ($M_{16}$)				& 	1.912	& 	14.479	& 	41.832		& 	82.289		& 	161.973		& 	307.109		&  	594.187	\\
				\hline
				Page fault handling in kernel space ($M_{5}$)					& 	0.003	& 	0.3	&	1.68		& 	3.34		& 	8.39		& 	16.79		&  	33.58	\\
				\hline
				Page fault handling in userspace ($M_{6}$)					& 	2.5	& 	27.3	&	152.3		& 	347.1		& 	882.8		& 	1,585		&  	3,483	\\
				\hline
				Hypercalls disable\_logging ($M_{14}$)								& 	0.042	& 	0.047	& 	0.138		& 	0.156		& 	0.189		& 	0.203		&  	0.208	\\
				\hline
				Ring buffer	copy ($M_{18}$)								& 	0.003	& 	0.01	&	0.03		& 	0.048		& 	0.109		& 	0.383		&  	0.671	\\
				\hline
				Reverse mapping ($M_{17}$) 								& 	6.183	& 	24.653	& 	85.117		& 	255.437		& 	1,211	& 	4,123	&  	15,738\\
				\hline
			\end{tabular}
	\caption{Cost of internal metrics which are influenced by Tracked memory size.}
	\label{tab:basic-metrics-2}		
\end{table*}

\begin{figure}[!h]
	\centering 
	\includegraphics[width=.8\columnwidth]{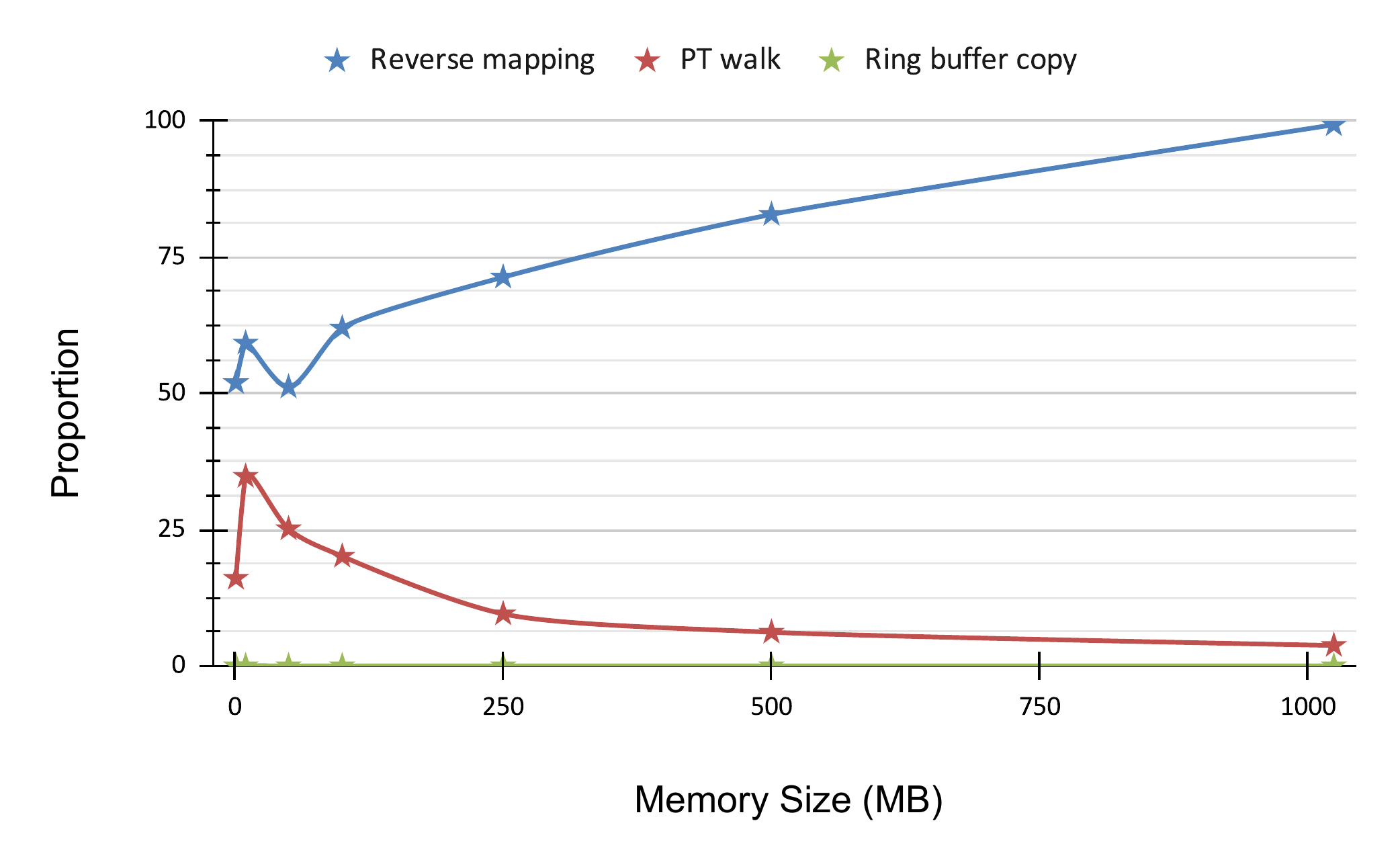}%
	\caption{SPML bottleneck.
	Reverse mapping appears to be the bottleneck.
	We used the micro-benchmark to realise this experiment.}
	\label{fig:spml-bottleneck}
\end{figure}
\subsection{Besic costs}
\label{sec:basic-costs}

\paragraph{Methodology}
To evaluate \texttt{/proc}, \texttt{userfaultfd}, and SPML we performed experiments on the real machine.

Because of the lack of an EPML capable machine, we build equation~\ref{eq:formula-2} which approximates the execution time of a given benchmark under EPML.
To explain that equation, let us reuse the same nomenclature (Tracker and Tracked) introduced in \Cref{sec:motivations}.
Notice that we are interested in evaluating the performance of Tracker and Tracked.
We use the term entity to refer to one or the other.
Given $P_{ideal}$ the ideal execution time of an entity, $C_{vmread}$ and $C_{vmwrite}$ respectively the costs of \texttt{vmread} and \texttt{vmwrite}, $N$ the sum of the number of scheduled-in and scheduled-out events which involves Tracked, and $C_{copyrb}$ the cost of copying the ring buffer content, then the performance of the entity under EPML (noted $P_{EPML}$) is estimated using this equation:
\begin{equation}
P_{EPML} = P_{vanilla} + N \times (3 \times C_{vmwrite} + C_{vmread}) + C_{copyrb}
\label{eq:formula-2}
\end{equation}

All the resutls presented in this section are mean of 5 runs. 

We present in this section the cost of all internal metrics which allow to understand higher level performance metrics.
The value of these metrics also tell us about the scalability of each tracking technique.
The first column of Table~\ref{tab:basic-metrics-1} lists the metrics, organized in nine categories.
For each metric, we indicate if its value is influenced or not by Tracked memory size (second column of table~\ref{tab:basic-metrics-1}).
For metrics which are agnostic to Tracked memory size, their basic cost is reported in the third column of table~\ref{tab:basic-metrics-1}.
For the other metrics, we reported their basic costs in table~\ref{tab:basic-metrics-2} while varying Tracked memory size.
We also indicate in column four of table~\ref{tab:basic-metrics-1} which tracking methods influence the metric.

\begin{table*}[h]
	\scriptsize
	\centering
   \begin{tabular}{ |l|c|c|c|c| }
		\hline
		& \texttt{/proc} & \texttt{userfaultfd} & SPML & EPML \\ 
		\toprule
		\hline
		influenced metrics & $M_1$, $M_5$, $M_{15}$, $M_{16}$ & $M_1$, $M_2$, $M_5$, $M_6$ & $M_1$, $M_3$, $M_4$, $M_9$, $M_{11}$, & $M_1$, $M_3$, $M_4$, $M_7$,\\
		 &  &  & $M_{13}$, $M_{14}$, $M_{16}$, $M_{17}$, $M_{18}$ &  $M_8$, $M_{10}$, $M_{12}$,  $M_{18}$ \\
		\hline
		metrics depending on Tracked mem. size & 3 ($M_5$, $M_{15}$, $M_{16}$) & 3 ($M_2$, $M_{5}$, $M_{6}$) & 4 ($M_{14}$, $M_{16}$, $M_{17}$, $M_{18}$)  & 1 ($M_{18}$) \\
		\hline
		metrics involved in the monitoring phase & 1 ($M_5$) & 2 ($M_5$, $M_6$) & 2 ($M_{13}$, $M_{14}$)  & 2 ($M_7$, $M_8$) \\
		\hline
		the two most costly metrics & $M_{5}$, $M_{16}$ & $M_{5}$, $M_{6}$  & $M_{16}$, $M_{17}$ & $M_{10}$, $M_{12}$ \\
		\hline		
		metrics which impact scalability & \multirow{2}{*}{\textbf{3} ($M_5$, $M_{15}$, $M_{16}$)}  & \multirow{2}{*}{\textbf{3} ($M_2$, $M_5$, $M_6$)} & \multirow{2}{*}{\textbf{4} ($M_{14}$, $M_{16}$, $M_{17}$, $M_{18}$)} & \multirow{2}{*}{\textbf{1} ($M_{18}$)} \\		
		from Tracker point of view		 &  &  &  &  \\	
		\hline
		metrics which impact scalability & \multirow{2}{*}{\textbf{3} ($M_5$, $M_{15}$, $M_{16}$)} & \multirow{2}{*}{\textbf{2} ($M_5$, $M_6$)} & \multirow{2}{*}{\textbf{2} ($M_{13}$, $M_{14}$)} & \multirow{2}{*}{\textbf{2} ($M_7$, $M_8$)}  \\		
		from Tracked point of view		 &  &  &  &  \\			
		\hline				
	\end{tabular}
	\caption{Tne influence of \texttt{/proc}, \texttt{userfaultfd}, SPML and EPML on internal metrics. $M_i$ are defined in table~\ref{tab:basic-metrics-1}.}
	\label{tab:analyse}    
\end{table*}

Table~\ref{tab:analyse} summarizes our analysis of the values reported in table~\ref{tab:basic-metrics-1} and table~\ref{tab:basic-metrics-2}:

\begin{itemize}
	\item \texttt{/proc}: it influences 4 metrics, including 3 which depend on Tracked memory size. 
	Among them, the most costly metric is page table walk in userspace ($M_{16}$) that can take up to $594 ms$.
	Page fault handling in kernel space ($M_{5}$) is the second most costly metric, that can take up to $33 ms$, which is quite significant in respect with the fact that it may involve frequently during the monitoring phase.
	This is why it dramatically impacts \texttt{/proc} scalability (from both Tracked and Tracker points of view).
	
	\item \texttt{userfault}: it influences 3 metrics, including 2 which depend on Tracked memory size.
	The most costly metric is page fault handling in userspace ($M_6$), which costs up to $3,483 ms$ when Tracked memory size is 1GB.
	This metric, in addition, is involved during the monitoring phase, thus impacting the scalability of \texttt{userfaultfd} (from both Tracked and Tracker points of view).
	
	\item SPML: it influences 10 metrics, including 4 which depend on Tracked memory size.
	The most costly metric is reverse mapping ($M_{17}$) which takes up to $15K ms$ for 1GB Tracked memory size.
	This metric will only impact the scalability of Tracker because it is not involed in the monitoring phase.
	Fig.~\ref{fig:spml-bottleneck} presents the proportion of time taken by each step of the collection phase in SPML. 
	We can observe that reverse mapping is definitely the bottleneck of SPML as it represents in average more than $60\%$ of the total collection time.	
	
	\item EPML: it influences 8 metrics, including only one metric depends on Tracked memory size.
	The most costly metric is PML initialization ($M_{10}$) which also includes VMCS shadowing initialization.
	It costs about $5,878 ms$.
	Because this metric does not depend on Tracked memory, it does not impact the scalability of Tracker. 
	The metrics that impact the scalability of Tracked are \texttt{vmread} ($M_{7}$) and \texttt{vmwrite} ($M_{8}$) whose costs are very low (less than $1 \mu s$). 
	This makes EPML scalable.
\end{itemize}

\subsection{Impact on Tracked}
\label{sec:microbench-results}
\begin{figure}[!h]
	\centering 
	\includegraphics[width=.8\columnwidth]{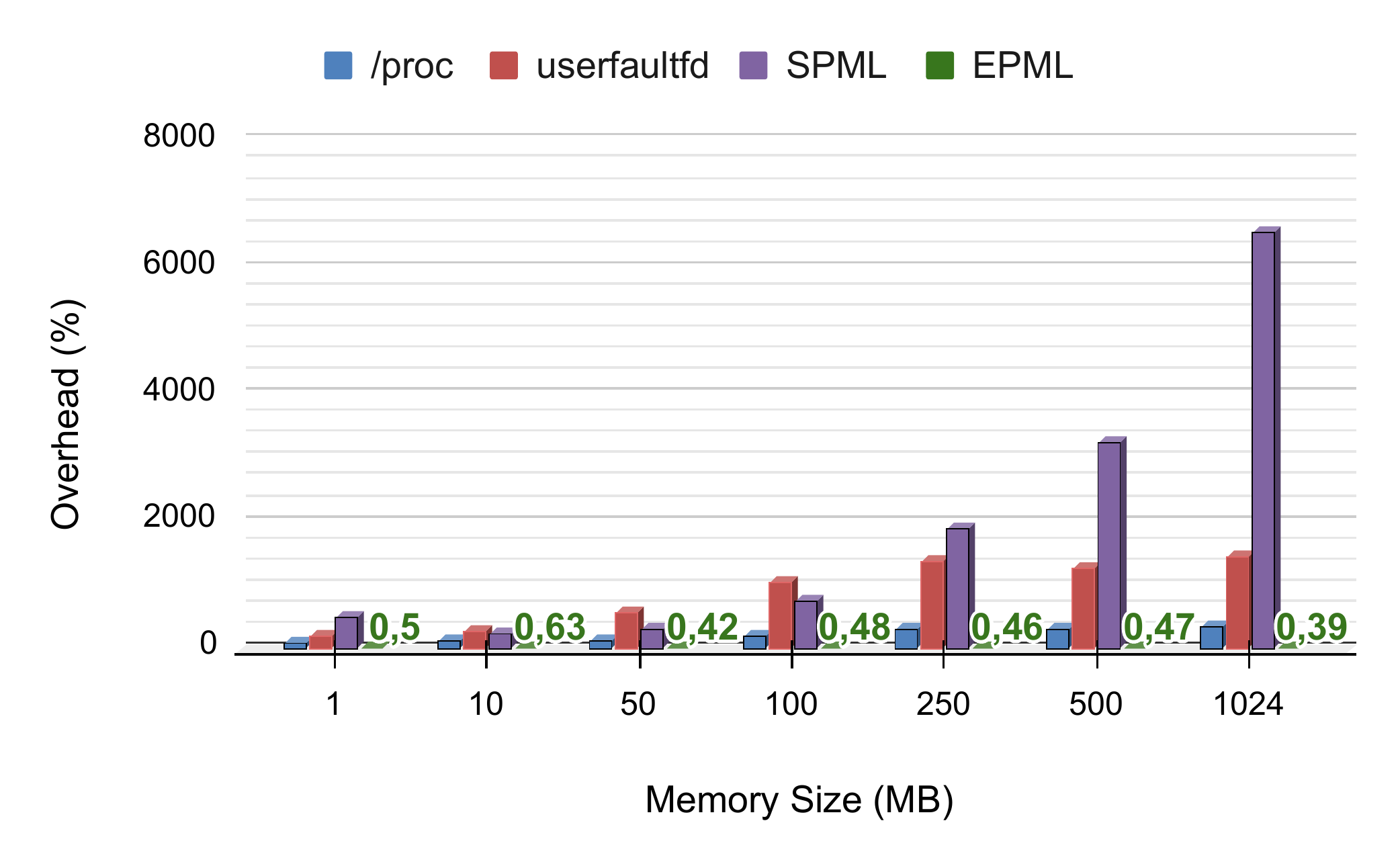}%
	\caption{Overhead of each tracking technique on the micro-benchmark.
	The values for EPML shown as they are too low.}
	\label{fig:microbench-overheads}
\end{figure}

This section presents the impact of each dirty page tracking technique on Tracked.
For these evaluations, Tracked is the micro-benchmark. 
Fig.~\ref{fig:microbench-overheads} presents the results.
We observe that, for almost all memory sizes, SPML incurs the greatest overhead compared to the other techniques. 
The overhead of EPML is almost null regardless Tracked memory size. 
This makes EPML the best technique amongst all and confirms its scalability.

%---------------------------evals-criu---------------------
\subsection{CRIU}
\label{criu-results}
We compare three prototypes:
CRIU \texttt{/proc},
CRIU SPML and,
CRIU EPML.
We did not consider it useful to implement a \texttt{userfaultfd}-based CRIU version because our main goal is to outperform CRIU \texttt{/proc}, which is the default implementation of CRIU.
Notice that, according to our nomenclature, CRIU is Tracker.

\begin{figure}[!h]
    \centering
    \begin{subfigure}[b]{.35\textwidth}
        \centering
        \includegraphics[width=1\columnwidth]{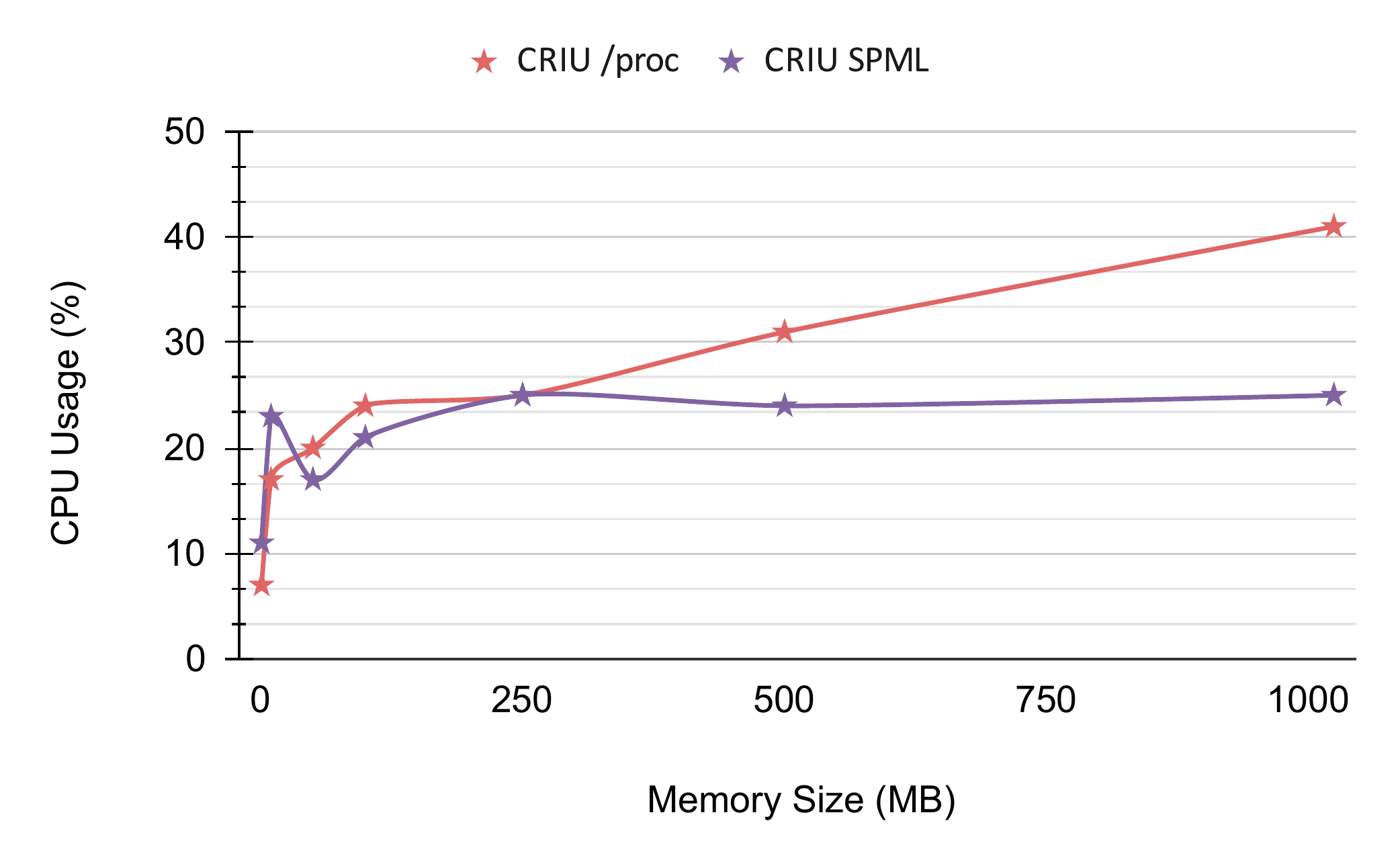}%
        \subcaption{During the initialization phase}
    \label{fig:criu-cpu-a}        
    \end{subfigure}
    \begin{subfigure}[b]{.35\textwidth}
        \includegraphics[width=1\columnwidth]{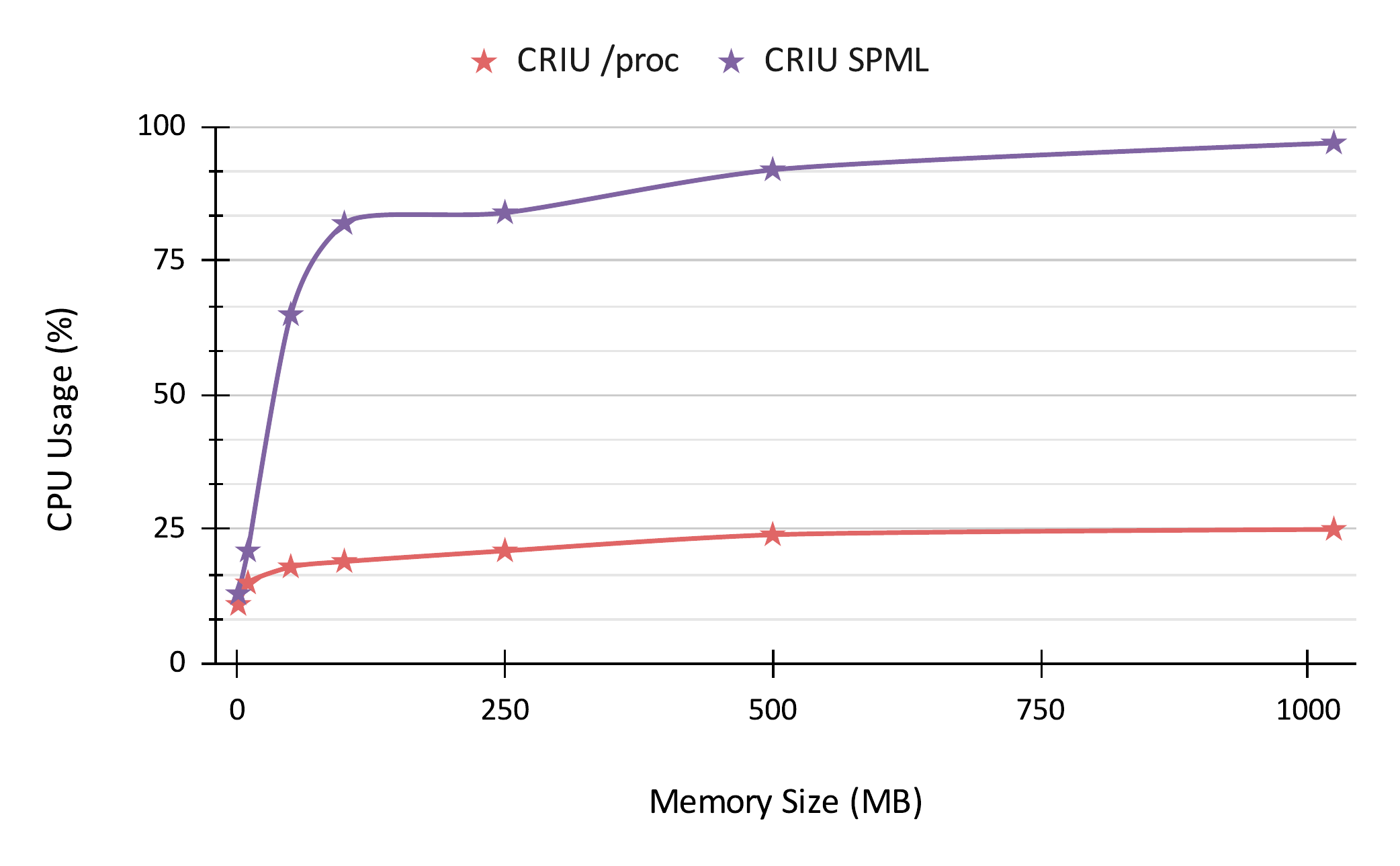}%
        \subcaption{During the collection and exploitation.}
    \label{fig:criu-cpu-b}        
    \end{subfigure}
    \caption{CPU usage generated by CRIU \texttt{/proc} and CRIU SPML}
    \label{fig:criu-cpu}
\end{figure}
\paragraph{Impact on CRIU}
We use the micro-benchmark (Tracked) for this evaluation.
We evaluate the impact on each CRIU phase.
First, Fig.~\ref{fig:criu-cpu} shows CPU consumption values for CRIU \texttt{/proc} and CRIU SPML for the initalization (Fig.~\ref{fig:criu-cpu-a}), collection+exploitation\footnote{In CRIU these two phases are combined.} (Fig.~\ref{fig:criu-cpu-b}) phases.
In CRIU, collection+exploitation is the actual checkpointing phase, which is the most critical one.
Compared to \texttt{/proc}, which requires to clear dirty bits at initialization time, we can observe that SPML reduces the CPU usage during by about $14.3\%$ in average.
Also, the CPU usage with CRIU SPML is almost constant while it increases with Tracked memory size in CRIU \texttt{/proc}. 
We did not collect CPU consumption values for CRIU EPML according to the fact that it is emulated.
Nevermind, for the initialization phase we think that it would lead to the same results as CRIU SPML.
Concerning the collection+exploitation phase, its CPU consumption would be lower than both CRIU SPML and CRIU \texttt{/proc}.
It would correspond to CPU usage to write dirty pages to disk.

Table~\ref{tab:times-criu} shows the checkpointing time.
We can see that CRIU EPML is the best prototype.
It outperforms CRIU \texttt{/proc} and CRIU SPML by about 61\% and 1,515\% respectively, when Tracked memory size is 1GB.

\begin{table}[!htb]
	\scriptsize
	\centering
		\begin{tabular}{ |l|c|c|c|c|c|c|c| } 
			\hline
			Size & 1MB & 10MB & 50MB & 100MB & 250MB & 500MB & 1GB \\ 
			\toprule
			\hline
			\texttt{/proc} & 107.17 & 132.45 & 280.35 & 399.26 & 577.97 & 889.47 & 1,627 \\
			\hline
			SPML & 112.81 & 148.66 & 327.42 & 756.09 & 2,123 & 5,520 & 16,326 \\			
			\hline
			EPML & 105.33 & 119.36 & 237.74 & 327.96 & 405.24 & 607.00 & 1,011 \\ 
			\hline			
		\end{tabular}
	\caption{CRIU checkpointing time with CRIU \texttt{/proc}, CRIU SPML and, CRIU EPML.
	The tracked application is the micro-benchmark.
	All the values are given in milliseconds (ms).}
	\label{tab:times-criu}		
\end{table}

\begin{figure}[!h]
	\centering 
	\includegraphics[width=.8\columnwidth]{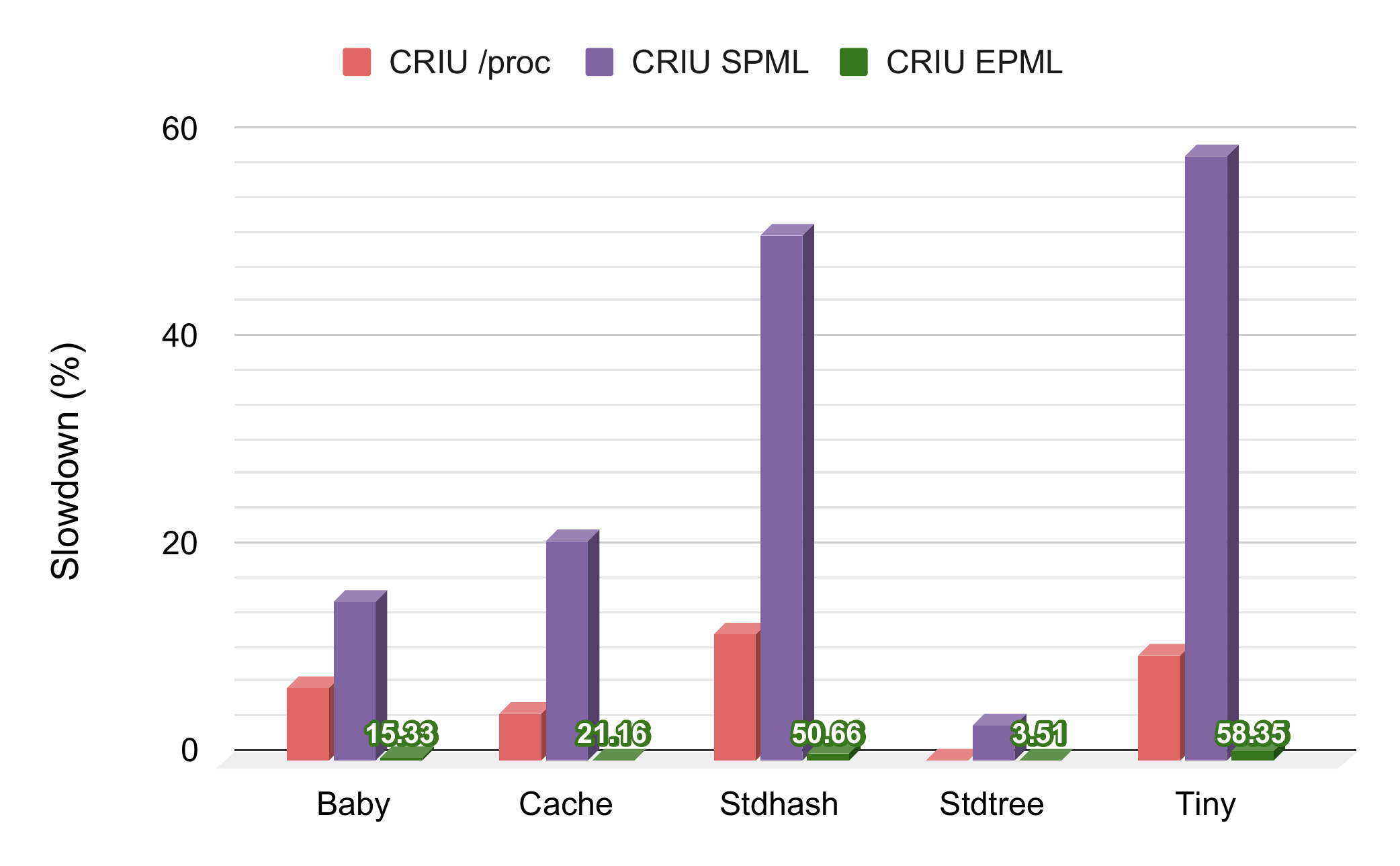}%
	\caption{Overhead of CRIU \texttt{/proc}, CRIU SPML, CRIU EPML on five tkrzw engines.
	Lower is best.}
	\label{fig:criu-macrobench}
\end{figure}

\begin{table}[h]
	\scriptsize
	\centering
		\begin{tabular}{ |c|c|c|c|c|c| }
			\hline
			Baby 		& Cache 	& Stdhash 	& Stdtree 	& Tiny \\ 
			\toprule
			\hline
			833MB 		& 596MB 		& 2.4GB 		& 2.4MB		& 2.2GB \\        
			\hline
		\end{tabular}	
	\caption{Memory consumption of tkrzw engines.}
	\label{tab:tkrzw-sizes}    
\end{table}
\paragraph{Impact on Tracked}
For this evaluation Tracked is a tkrzw engine.
Table~\ref{tab:tkrzw-sizes} shows the amount of memory consumed by each engine during the experiment.
We use these memory consumption values to compute, in the case of EPML, the cost of the internal metrics that depend on Tracked memory size.
Fig.~\ref{fig:criu-macrobench} presents the overhead of each solution.
We can observe that SPML significantly slowdowns the tracked application.
An overhead of up to 58\% is observed for Tiny.
The performance difference between CRIU SPML and CRIU \texttt{proc} for Tiny is 114\%. 
Concerning CRIU EPML, it leads to the lowest overhead.
It ameliorates CRIU \texttt{/proc} by about 11\% on Stdhash.

\subsection{Potential Missed Addresses under SPML}
\label{sec:missed-pages}
\begin{figure}[!h]
	\centering 
	\includegraphics[width=.6\columnwidth]{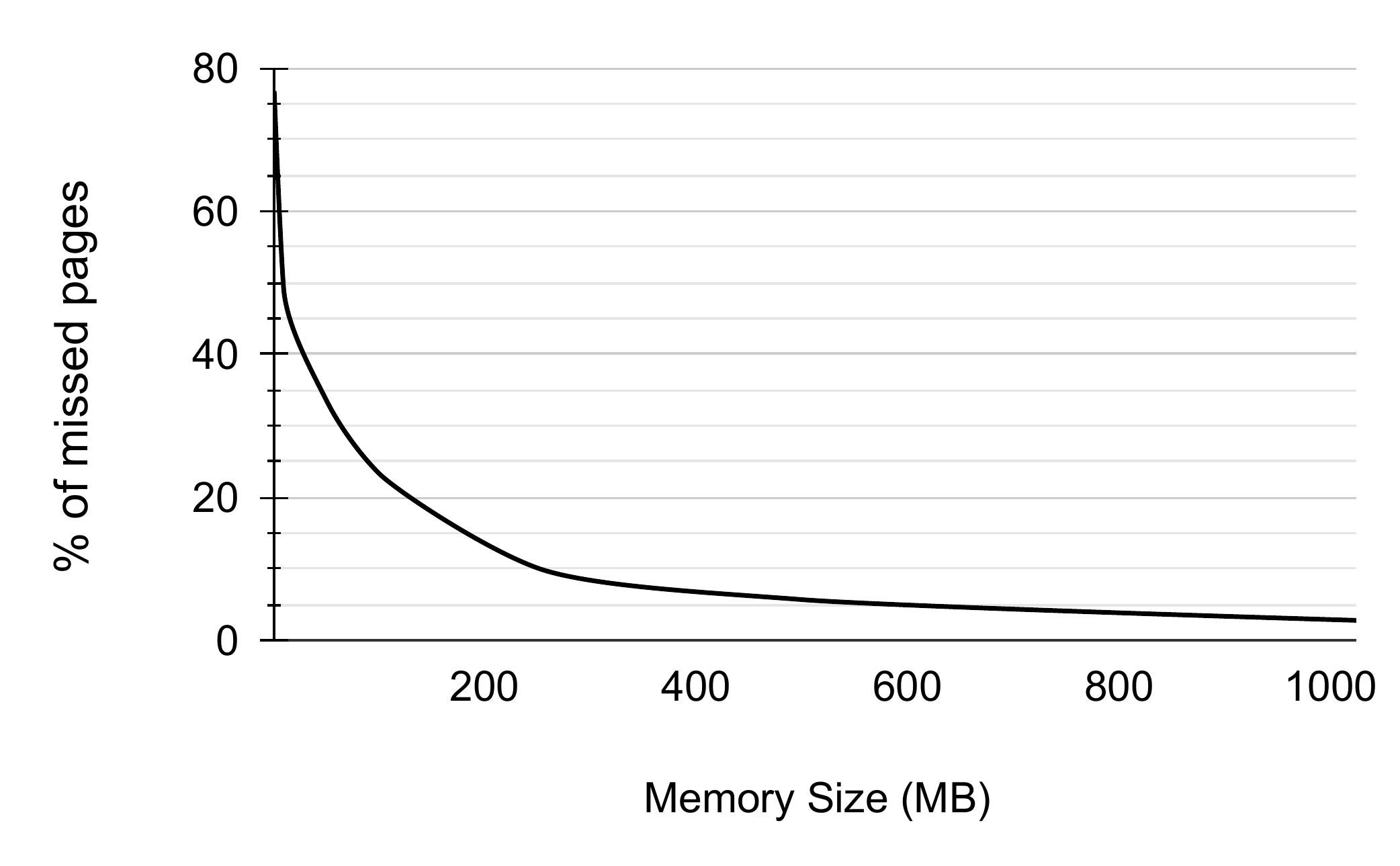}%
	\caption{Proportion of missed pages with SPML.}
	\label{fig:missed-pages}
\end{figure}

As stated in \Cref{sec:shadow-pml-limit}, SPML may miss some dirty page addresses during reverse mapping.
This section evaluates the proportion of these addresses while varying the dirty page working set size of Tracked.
The latter is the micro-benchmark.
Fig.~\ref{fig:missed-pages} shows the results.
We can observe that the proportion decreases with the working set size: from $76.7\%$ down to $2.7\%$.
Therefore, applications with a small dirty page working set size will suffer a lot from this SPML limitation. 
Anyway, losing even one dirty page could be problematic for the system to which dirty page tracking are destinated.
For example, the fault recovery system which relies on the latest checkpoint file will try to restart the application with an inconsistent state, which may crash.
This may compromise the high availability goal that checkpoint-restore was supposed to bring.
By design, EPML does not suffer from this problem.

\subsection{Coexistence with the Hypervisor}
\label{sec:coexistence-eval}
The purpose of this section is to validate that both the VM and the hypervisor can use PML simultaneously.
To this end, we realize the following experiment.
We start two VMs noted $VM_1$ and $VM_2$.
In $VM_1$ we launch our micro-benchmark along with a tracking system.
At the same time, we launch the live migration of $VM_2$. 
Recall that Xen live migration uses PML when it is supported by the machine, which is our case.
The results are as follows.
First, we observed that the hypervisor was able to complete live migration without crashing.
Second, we observed that the migration time increases by about $45\%$ compared to when it is performed alone.
This is explained by the multiplication of the number of vmexit events due to PML buffer full.
Notice that the same buffer is shared among the hypervisor and VMs.
By design EPML does not include this overhead as it allows the hypervisor and the guest evolve independently on their dedicated PML buffer.
%---------------------------evals-criu---------------------

%---------------------------evals---------------------

%---------------------------rw---------------------
\section{Related work}
\label{rw}
\paragraph{Nested virtualization}

The main source of performance degradation in virtualized environements are VM traps.
The latter lead to the suspension of the execution of the VM and also to cache pollution~\cite{266847} dues to context switch.
The number of VM traps increases at least by a factor of two in nested virtualized environements~\cite{vilanova.isca.2019}.
The reduction of VM traps is a hot topic in both non-nested~\cite{10.1145/2150976.2151020,10.1145/3381052.3381317,266847} and nested virtualized systems~\cite{lim.dvh.asplos.2020,vilanova.isca.2019,neve.sosp.2017}.

Device passthrough is a simple approach for improving I/O performance in nested and non-nested virtualized environments by providing a direct access to the VM.
However, it dedicates the entire device to a single VM, resulting to sub-optimal resource utilization.
In addition, device passthrough does not permit VM live migration, which is an important operation for cloud providers as it is used for maintenance.
VMCS shadowing~\cite{vmcs-shadowing}, that EPML leverages, has been introduced by Intel to reduce traps when a nested hypervisor accesses some VMCS fields.
$SV_T$, by Vilanova et al.~\cite{vilanova.isca.2019}, exploited simultaneous multithreading (SMT) processors to minimize VM traps.
$SV_T$ runs every nested virtualization level on a separate SMT thread and it replaces VM trap and VM resume to avoid context switches between nested hypervisors and the host hypervisor (the one which directly runs atop the hardware).
In $SV_T$, only one SMT thread can run at a given time leading to core waste.

DVH, by Lim et al.~\cite{lim.dvh.asplos.2020}, proposed that the host hypervisor to provide virtual devices directly to nested VMs without the intervention of intermediate hypervisors.
The intermediate hypervisors only intervers at virtual device initialization time to make it visible and directly accessible to the nested VM.
The authors illustrated DVH with four devices: virtual-IO, virtual timer, virtual IPI and virtual idle.
Although DVH is promising, its application to all devices which composed a full hardware is unpractical.
With OoH, we are advocating for exposing only hardware virtualization features that could help applications, which is tractable.

\paragraph{Dirty page tracking}
This activity is necessary for both hypervisors and processes.
The hypervisor relies on it to perform pre-copy based live migration and also checkpointing.
Dirty page monitoring is at the heart of concurrent garbage collectors and other userspace processes such as CRIU for container and process checkpointing, or Redis for dumping the database.
So far, the main approach used for monitoring dirty pages is two steps: invalidation of PTE dirty bit and present bit, and page-fault interception.
To minimize the overhead of this approach, some alternatives have been proposed.
For the hypervisor, Intel introduced PML, the hardware feature that we study in the present paper.
Bitchebe et al.~\cite{10.1145/3453933.3454018} showed that PML can decrease both VM live migration and checkpointing duration.
The authors also extended PML to also log read pages in order to efficiently estimate VM working set size.
In non-virtualized environments, Lu et al.~\cite{Lu:2017:FPM:3050748.3050751} built a memory allocator which maps several objects to the same physical page, thus reducing the number of tracked pages.
Althougt This solution does not avoid frequent interruption of the tracked process due to page faults.
%---------------------------rw---------------------

%---------------------------conclusion---------------------
\section{Conclusion}
\label{conclusion}
In this paper we introduced Out of Hypervisor (OoH), a new research axis that advocates the exposition of individually current hypervisor-oriented hardware virtualization features to the guest OS so that its processes could also take benefit from those features.
We illustrated OoH with Intel PML (noted OoH-PML), a feature which allows efficient dirty page tracking.
We prototyped OoH-PML following two designs namely Shadow (SPM)L and Extended PML (EPML).
The former requires no hardware changes but incurs significant performance overhead.
It is not the case of EPML which extends the original PML to avoid SPML limitations.
We evaluated and compared SPML and EPML with two popular dirty page tracking techniques namely \texttt{/proc} and \texttt{userfaultfd}.
We considered CRIU as the usecase where the target application is tkrzw, a database engine.
The evaluation results showed that the different techniques can be classified as follows:
SPML, \texttt{userfaultfd}, \texttt{/proc}, and EPML.
We are currently applying OoH to Intel SPP in order to improve secured heap memory allocators.
%---------------------------conclusion---------------------

%-------------------------------------------------------------------------------
\bibliographystyle{IEEEtranS}
\bibliography{main}
%%%%%%%%%%%%%%%%%%%%%%%%%%%%%%%%%%%%%%%%%%%%%%%%%%%%%%%%%%%%%%%%%%%%%%%%%%%%%%%%
\end{document}